\documentclass[12pt,preprint]{aastex}
\usepackage{graphicx}
\usepackage{color}
\usepackage{natbib}

\newcommand{\chandra}{{\it Chandra}}
\newcommand{\aox}{\ifmmode \alpha_{ox}\else$\alpha_{ox}$\fi}
\newcommand{\mone}{\ifmmode ^{-1}\else$^{-1}$\fi}
\newcommand{\mtwo}{\ifmmode ^{-2}\else$^{-2}$\fi}
\newcommand{\degs}{\ifmmode ^{\circ}\else$^{\circ}$\fi}
\newcommand{\mv}{\ifmmode {m_{V}}\else${m_{V}}$\fi}

\newcommand{\gae}{\mathrel{\raise .4ex\hbox{\rlap{$>$}\lower 1.2ex\hbox{$\sim$}
} }}
\begin{document}
\shorttitle{Intermittent activity in AGN}

\slugcomment{\today}


\title{Accretion Disk Model of Short-Timescale Intermittent Activity in Young Radio Sources}

\author{Bo\.zena Czerny$^{1}$, Aneta Siemiginowska$^{2}$, Agnieszka Janiuk$^{1}$, B{\l}a\.zej 
Nikiel-Wroczy\'nski$^{1, \, 3}$, {\L}ukasz Stawarz$^{4, \, 3}$}

\affil{$^{1}$Nicolaus Copernicus Astronomical Center, ul. Bartycka 18, 00-716 Warsaw, Poland}
\affil{$^{2}$Harvard Smithsonian Center for Astrophysics, 60 Garden St, Cambridge, MA 02138, USA}
\affil{$^{3}$Astronomical Observatory, Jagiellonian University, Orla 171, 30-244 Cracow, Poland}
\affil{$^{4}$Kavli Institute for Particle Astrophysics and Cosmology, Stanford University, Stanford, CA 94305, USA}

\email{bcz@camk.edu.pl}


\label{firstpage}

\begin{abstract}

We associate the existence of short-lived compact radio sources with
the intermittent activity of the central engine caused by a radiation
pressure instability within an accretion disk.  Such objects may
constitute a numerous sub-class of Giga-Hertz Peaked Spectrum sources,
in accordance with the population studies of radio-loud active
galaxies, as well as detailed investigations of their radio
morphologies.  We perform the model computations assuming the
viscosity parametrization as proportional to a geometrical mean of
the total and gas pressure. The implied timescales are consistent with
the observed ages of the sources. The duration of an active phase for
a moderate accretion rate is short enough ($< 10^3-10^4$ years) that
the ejecta are confined within the host galaxy and thus these sources
cannot evolve into large size radio galaxies unless they are close to
the Eddington limit.

\end{abstract}


\section{Introduction}

\label{sec:intro}

The duty cycle of active galactic nuclei is still a subject of 
debate. Different mechanisms are likely to operate. On long
timescales, up to $10^8 $ yrs, the merger activity is broadly
considered as the cause of an enhanced accretion flow and the main
provider of material for the black hole growth. However, at later
stages, active galactic nuclei are also likely to pass through
subsequent stages of higher and lower activity. Intermittent behavior
of this type is likely to be caused by minor mergers or instabilities
in the accretion flow.

There are several observational arguments for this intermittent
behavior. For example, morphology of X-ray clusters studied with 
\chandra\, X-ray Observatory and {\it XMM-Newton} indicate a past 
intermittent AGN activity of the central dominant galaxy that is 
responsible for heating of the central cluster regions. X-ray images 
show cavities and ripples that give the timescales of repetitive 
outbursts and provide a measure of their power (e.g. Ruszkowski 
et al. 2004; McNamara \& Nulsen 2007; Sanders \& Fabian 2008). Such AGN 
intermittency seems to be closely related to the feedback process
operating in galaxy clusters.

Radio galaxies that are not associated with clusters, also provide
signatures of the intermittent activity.  First, in many cases studies
of radio structure directly showed evidence of two or more of active
periods (e.g. Burns et al. 1983; Bridle et al. 1989; Clarke et
al. 1992; Schoenmakers et al. 2000a,b; Marecki et al. 2003; Jamrozy et
al. 2007). In some sources the new radio structure is highly inclined
to the old one (e.g. Cheung 2007), thus indicating an independent
accretion/black hole merger episode resulting in renewing activity of
the central engine accompanied by the re-orientation of the jet axis
(e.g. Merritt
\& Ekers 2002; Gopal-Krishna et al. 2003; Blundell 2008; Lal et al. 2008). 
Some sources of this type are very compact, only a few kpc in size
(Marecki et al. 2006). In other sources, both structures (old and new)
are aligned (e.g. Saripalli et al. 1996; Schoenmakers et al.  2000a,b;
Konar et al. 2006; Marecki et al. 2006), which is consistent with the
development of perturbations in the accretion rate within the existing
accretion disks and subsequent intermittent radio activity along 
the same (unchanged) direction of the outflow 
(Siemiginowska et al. 1996; Siemiginowska \& Elvis 1997; Hatziminaoglou
et al. 2001; Janiuk et al. 2004). An extreme example of such an activity
is a triple-double source in which three periods of activity led to
the aligned features (B0925+420; Brocksopp et al. 2007).

Second, the population studies show the existence of far too many
compact (``young'') sources in comparison with the number of galaxies
with extended old radio structures (O'Dea \& Baum 1997; Snellen et
al. 2000) although the full discussion is complex due to various
limitations of the available samples (Tinti \& De Zotti 2006). If the
total activity period lasts $10^8$\,yrs, the number of sources with
the ages below $10^3$\,yrs should be roughly $10^5$ times lower than
the number of sources older than $10^{3}$\,yrs, which is not the
case. One possible explanation is that the jet gets frustrated, and
the source is in fact old, although it is small in size (van Breugel
et al. 1984, Readhead et al. 1996) . However, the important argument
against this interpretation is the lack of the required amount of the
thermal matter to quench the jet expansion within host galaxies of GPS
sources (Siemiginowska et al.  2005, and references therein). A single
short-lasting episode of activity can explain the observed ages
(Readhead et al. 1996) but there is no natural explanation of such a
timescale. Therefore, a possible explanation is that the activity is
intermittent and the source is reborn every few thousand years,
without displaying any strong evidence of the previous activity phase,
as discussed by several authors (e.g. Reynolds \& Begelman 1997;
Kaiser et al. 2000; Kunert-Bajraszewska et al. 2005, 2006; Jamrozy et
al. 2007; Giroletti 2008). Detection of several candidates for dying
compact sources supports this view (e.g. Giroletti et al. 2005;
Kunert-Bajraszewska et al. 2006; Parma et al. 2007).

An accretion disk scenario is a likely explanation of the intermittent
behavior on timescales much shorter than $10^8$\,yrs. Studies of
accretion disks in other accreting objects show that cold accretion
disks are subject to two types of instabilities: ionization
instability, operating on longer timescales and responsible for dwarf 
nova and X-ray nova phenomenon (Osaki 1974; Meyer \& Meyer-Hofmeister 
1981; Smak 1982; see Lasota 2001 for a review), and radiation pressure
instability, operating on shorter timescales and responsible for regular 
outbursts in GRS 1915+105 (Nayakshin et al. 2000; Janiuk et al. 2000, 
2002; Merloni \& Nayakshin 2006; see also Fender \& Belloni 2004, and 
Done et al. 2004 for general reviews on this source). The
ionization instability in the context of AGN was studied by a number
of authors (Lin \& Shields 1986; Clarke 1988; Mineshige \& Shields
1990; Siemiginowska et al. 1996; Siemiginowska \& Elvis 1997;
Hatziminaoglou et al. 2001; Janiuk et al. 2004). The derived separations
between outbursts were on order of $10^6$\,yrs for a $10^8\,M_{\odot}$
black hole, while the outburst duration was an order of magnitude
shorter.

In this paper we concentrate on the shorter timescale, radiation pressure 
driven instability operating above the threshold accretion rate $\dot m=0.025$
(in the Eddington units), and we test whether this mechanism is likely to be
responsible for very short average ages of Compact Symmetric Objects
(CSO), a subgroup of GigaHertz Peaked Sources (GPS).
In Section \ref{sec:data}, we present our sample of GPS sources.
In Section \ref{sec:model}, we present our model of the time evolution of 
the accretion disk under the radiation pressure instability, and  in Section 
\ref{sec:results} we show the resulting intermittent activity patterns for
a range of black hole mass, accretion rates and viscosity coefficients.
In Section \ref{sec:discussion} we discuss our results. 
Throughout this paper we use the cosmological parameters based on the
WMAP measurements  (Spergel et al. 2003):
H$_0=$71~km~sec$^{-1}$~Mpc$^{-1}$, $\Omega_M = 0.27$, and $\Omega_{\rm
vac} = 0.73$.

\section{Young Radio Sources: Timescales and Energetics}
\label{sec:data}

Giga-Hertz Peaked Spectrum (GPS) radio sources are compact and
selected based on the convex shape of the radio spectrum that peaks at
GHz frequencies (for a review see O'Dea 1998). They have a compact
radio morphology with the radio emission contained within $<1$\,kpc
(O'Dea 1998) of the host galaxy center. Their radio morphologies on
miliarcsec scales resemble the ones observed in large ($\gg 10$\,kpc)
radio galaxies with lobes, terminal hot spots and, in some cases,
clear core-jets structures. Because of this similarity and the
observed high radio power the GPS sources are thought to be the
precursors of large radio galaxies observed at the early stage of
their expansion (Fanti et al 1995; O'Dea \& Baum 1997; O'Dea
1998). However, as noted above, the number of these sources is too
high to support a uniform self-consistent expansion from a compact to
a large megaparsec scale radio source. In order to explain the
observed overabundance of the compact sources, Reynolds \& Begelman
(1997) suggested an intermittent activity of radio-loud active
galaxies, with the characteristic activity timescales of
$10^4-10^5$\,yrs based on the observed statistics (O'Dea \& Baum
1997).

Compact Symmetric Objects (CSO) comprise a subgroup of GPS sources that
was introduced as a separate class over 10 years ago (Wilkinson et
al. 1994) on the basis of their ``classical double'' but extremely compact 
($< 500$~pc) radio morphology. The small size of the radio
structures was usually interpreted as an indication of the age below
$10^4$ yrs (Readhead et al. 1996; Owsianik \& Conway 1998; Owsianik,
Conway \& Polatidis 1998; Taylor et al. 2000; for an alternative
suggestion of a frustrated jet see Carvalho 1994, 1998). The strong
observational support for the source's young age comes from VLBI
monitoring observations and measured hot spots propagation velocities
$\sim 0.1-0.2\,c$ indicating the kinematic ages of $\sim 10^3$\,yrs
(Conway 2002; Polatidis \& Conway 2003). It should be noted that there is 
no debate on the intrinsic compactness of CSOs, as opposed to many 
GPS objects displaying a clear core-jet morphology, since the latter
ones may \emph{appear} rather small due to the projection effects.


\begin{figure*}
\includegraphics[width=\columnwidth]{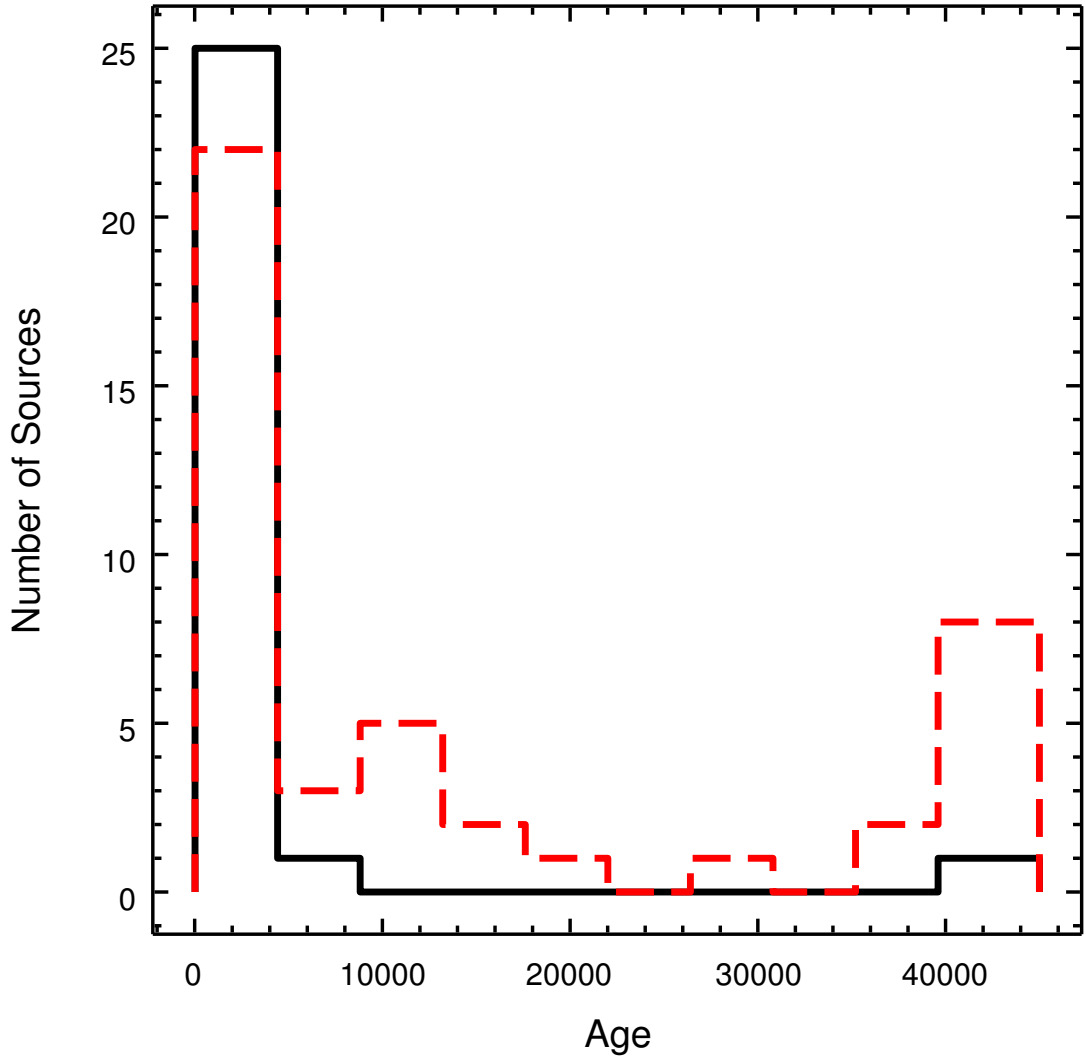}
\caption{The age distribution of the sources in our sample. 
The black solid line shows the distribution of the sources with the age
measured by the kinematic method and the red dashed line is the
distribution of the ages measured by the synchrotron method.}
\label{fig:histogram}
\end{figure*}


\begin{figure*}
\includegraphics[width=\columnwidth]{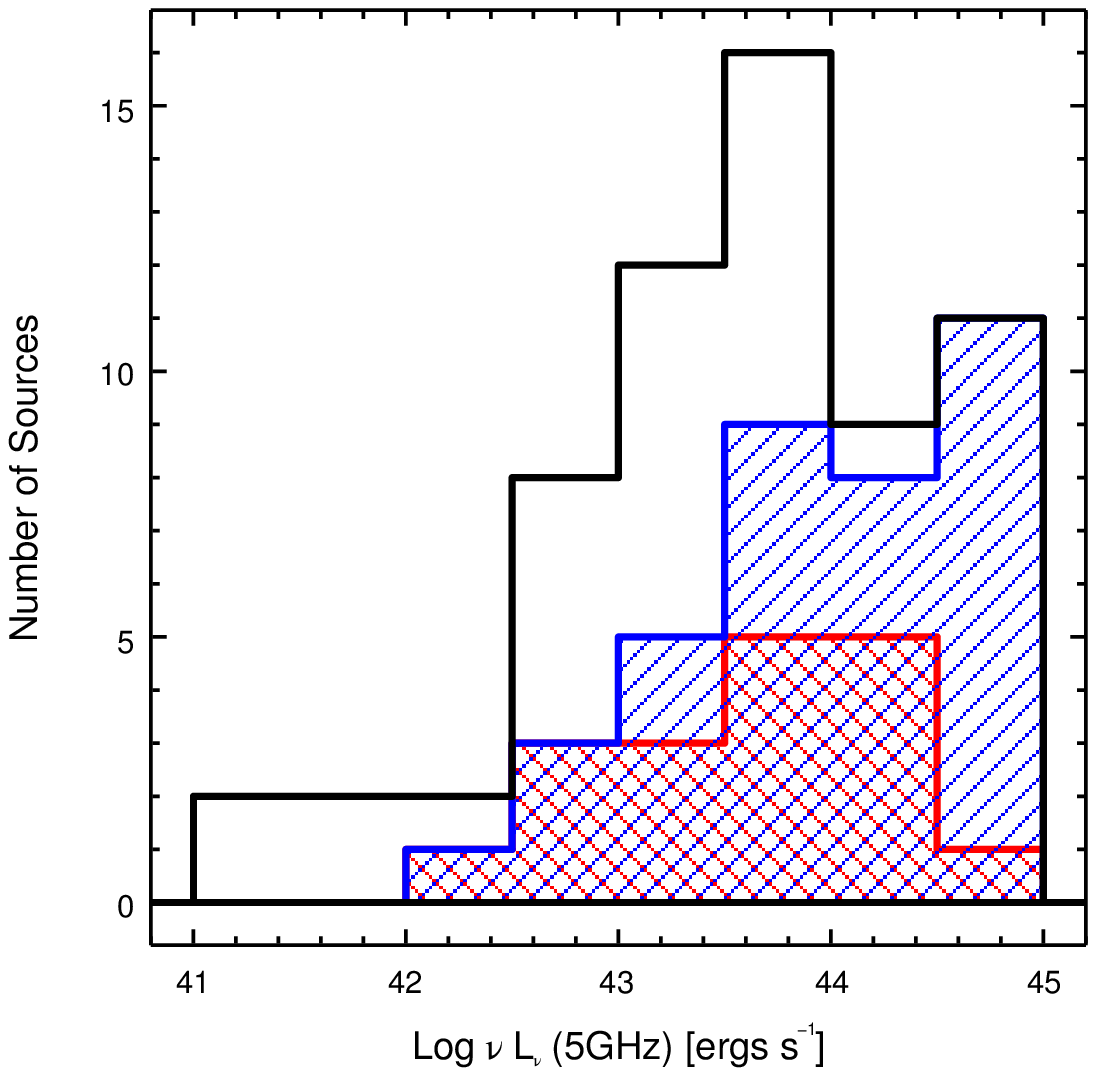}
\caption{The distribution of the 5GHz radio luminosity for the 72 sources in our sample 
(black continuous line, whole histogram). We also show the
contribution of different subclasses according to the source type:
type a (cross-shaded area), type b (shaded area) and the remaining
sources (unshaded area). For source classification see Table 1).}
\label{fig:lum_hist}
\end{figure*}

We compiled a sample of the GPS/CSO sources from the data available in
the literature. In this sample, 10 sources with a measured age are
from Polatidis \& Conway (2003) and 14 sources come from the studies
of the COINS sample by Gugliucci et al. (2005, 2007).  The kinematic
ages of these sources were determined on the basis of the measured hot
spot separation speed. 47 sources are taken from a much larger sample
of GPS sources (including larger size, $1-10$\,kpc Compact Steep
Spectrum, CSS) that were studied by Murgia et
al. (1999). Determination of their ages is based on the synchrotron
cooling time that gives a rather high upper limit. These sources had a
significant dispersion of the physical dimensions and derived ages.
Note that Murgia et al. (1999) distinguished two classes of sources:
the radio spectrum dominated by (a) lobes or (b) jets and hot
spots. The sources within the class ``a'' have the morphology similar
to the CSO sources and exhibit a correlation between the source age
and the radio linear size. Three individual sources were taken from
other publications (Orienti et al. 2007; Tschager et al. 2000).

The whole sample consists of 72 objects with a measured age, and the
details are provided in Table~\ref{tab1}. The majority of the age
values are between $200$\,yrs and $10^4$\,yrs, with three sources with
the spectral ages considerably longer than $10^5$\,yrs.  We plot the
age distribution in Fig.~\ref{fig:histogram}. The distribution looks
strikingly bimodal but this may well be an artifact of the sample
composition. Some age values (usually the shorter ones) are based on
the kinematic determination and thus represent a lower limit to the
source age, while the other age values (usually the longer ones) come
from the synchrotron aging method and give an upper limit to the true
age of the source (note that effects of radio source expansion, or
deviation from the equipartition may result in the true age of a
source to be shorter than its synchrotron age estimated for the fixed
radiative cooling rate).  A larger sample of sources with the ages
determined with a single method is necessary to test if any bimodality
is really present in the age distribution. On the other hand, we can
be reasonably convinced that the ages of the sources in our sample
capture the typical age values well.

We estimated the monochromatic radio luminosities ($\nu L_{\nu}$ at 5
GHz) for all the sources in our sample (see Table~\ref{tab1}) using
the measured 5~GHz flux density available in the
NED\footnote{http://nedwww.ipac.caltech.edu/index.html}
database. Figure~\ref{fig:lum_hist} shows the radio luminosity
distribution within the sample. These luminosities can be used to
infer the bolometric luminosities of the underlying accretion disks.
Note first that, assuming the ratio $L_{\rm syn}/L_{5\,{\rm GHz}}$ of
the order of one and no relativistic beaming (Wilkinson et al. 1994),
almost all the collected objects turn out to be characterized by
relatively large intrinsic synchrotron powers in a range $L_{\rm syn}
\sim 10^{43}-10^{46}$\,erg\,s$^{-1}$. This, with the $10\%$ radiative
efficiency found/claimed in the evolutionary models for young radio
sources (De Young 1993, Stawarz et al. 2008), implies the jet kinetic
luminosities in the range $L_{\rm j} \sim
10^{44}-10^{47}$\,erg\,s$^{-1}$, as required if GPS sources are indeed
precursors of the extended FR~I and FR~II radio galaxies.  Therefore,
the estimated jet powers set \emph{lower limits} on the accretion
rates in the considered objects which, with the standard $10\%$
radiative efficiency of the disk, would correspond to the accretion
luminosities $L_{\rm acc} > L_{\rm 5\,GHz}$.

On the other hand, if the observed radio flux is beamed, the intrinsic
synchrotron power of the analyzed sources may be smaller than quoted
above.  However, relativistic beaming is not expected to play any role
in the case of CSO objects (believed to be highly inclined), as well
as GPS sources with radio fluxes dominated by lobes (class
``a''). Only if the contribution of the inner jets (and possibly
hotspots) to the observed radio emission is significant, could the
assumed isotropy of the radio flux be questioned. And indeed, the
median $5$\,GHz luminosity of class ``b'' sources in our sample, $3
\times 10^{44}$\,erg\,s$^{-1}$, is one order of magnitude larger than
for the corresponding median for rest of the analyzed objects (see
Table~1). Such a difference is also apparent in
Figure~\ref{fig:lum_hist}, where in addition to the whole sample,
radio luminosity distribution is plotted for the two different classes
separately. Having said that, majority of the objects included in our
sample are of the CSO type or class ``a'', and thus our averaged
estimates presented in this section should be robust.

Meanwhile, five GPS/CSO sources studied with the XMM{\it -Newton} 
showed $2 - 10$\,keV luminosities comparable to $\nu L_{\nu}$ measured at 
$5$\,GHz (Vink et al. 2006). If the observed X-ray fluxes originate within the
accretion flow/disk corona, as advocated by several authors (e.g., 
Guainazzi et al. 2006, and references therein), the bolometric disk
luminosities are then expected to be larger than the observed X-ray 
ones by at least one order of magnitude (Koratkar \& Blaes 1999, Elvis et al. 1994).
We further note that the GPS/CSS sample of objects studied with {\it Chandra} 
revealed $2 - 10$\,keV fluxes by a factor of $\sim 5$ higher (on average) 
than $\nu L_{\nu}$ at 5 GHz (Siemiginowska et al. 2008), although the 
constructed broad-band spectral energy distributions showed also a large 
dispersion\footnote{Note in this context that 
in some radio-loud quasars the $2-10$\,keV luminosities are higher than 
the monochromatic $\nu L_{\nu}$ at 5 GHz flux by a factor of $\sim 300$ 
(Tengstrand et al 2008).}. Thus, the bolometric disk luminosities of GPS 
sources, including UV part, can easily be as large as $\sim 100 
\times L_{\rm 5\,GHz}$. The caution here is that the observed X-ray fluxes 
may be dominated by the inverse-Compton emission generated within compact 
lobes rather than by the accretion flow (Stawarz et al. 2008). Assuming 
an order of magnitude correcting factor of $100$, as a safe 
and conservative guess, implies the bolometric disk luminosities of the 
studied young radio sources $L_{\rm acc} \sim 10^{44} - 10^{47}$\,erg\,s$^{-1}$. 
If this correction term is much higher some of the sources may 
slightly exceed the Eddington luminosity, but this may be indeed acceptable 
for the brightest ones (see e.g. Collin et al. 2002).

Needless to say, the direct detection of the UV disk radiation in
GPS/CSO objects is not possible, since the accretion-related emission
at photon energies below $1$\,keV is known to be significantly
obscured. Note in this context that measurements of the HI absorption
in radio (Morganti et al. 2007, Pihlstr\"om et al. 2003) and the total
absorption in X-rays (Guainazzi et al. 2006, Vink et al 2006,
Siemiginowska et al. 2008, Tengstrand et al 2008) show significant
hydrogen column densities ($N_H \sim 10^{22}$\,cm$^{-2}$) in all the
studied GPS sources. In addition, the other independent estimates of
the bolometric luminosities on the basis of the ionization power do
not seem to work in young radio galaxies, as all the analyzed GPS
objects are underluminous in [OIII] line, as if the Narrow Line Region
is still in the formation process (Vink et al. 2006),  being in
addition possibly modified by the interaction with the expanding radio
lobes (Labiano 2008).

Only a few sources have the measurements of a black hole mass and we
give them in Table~2. The mass estimates are based on properties of
emission lines and the correlation established by the reverberation
mapping (Vestergaard \& Peterson 2006). These few measurements suggest
that the black holes in the sources of our sample cover a mass range
from $M \sim 10^8\,M_{\odot}$ to $M \sim 3 \times 10^9\,M_{\odot}$,
slightly higher than typical masses of radio sources in the Snellen et
al. (2003) sample ($\sim 10^8 M_{\odot}$). Also, for a given black
hole mass, the younger sources seem to have higher radio
luminosities. However, based on this small number of sources we cannot
formally assess any trends between the age and the radio
luminosity. Unfortunately, without black hole mass measurements for
majority of the studied objects, and with only roughly estimated
bolometric disk luminosities, we cannot directly calculate the
corresponding Eddington ratios. However, if we assume that the
representative value of the black hole mass in our sample is the
median from Table~2, i.e. $M \sim 3 \times 10^8\,M_{\odot}$, and take
the median monochromatic radio luminosity (for CSOs and class ``a''
sources only) from Table~1, namely $L_{\rm 5\,GHz} \sim 3 \times
10^{43}$\,erg\,s$^{-1}$, then with the anticipated correction factor
of $100$ one obtains the Eddington ratio on the order of $\dot{m} \sim
0.1$.

In addition to such average constraints, for those few sources with
the available masses of black holes we can estimate more precisely the
bolometric luminosities and the Eddington ratios, assuming the
observed fluxes given by Netzer et al (1996) and the template AGN
spectrum given by Elvis et al. (1994). The resulting luminosities and
accretion rate values are given in Table~2, and these agree with the
rough average estimates given above. Note that, most importantly, the
calculated accretion rates exceed the threshold for the radiation
pressure instability, ($\dot m=0.025$), for each analyzed source. The
bolometric corrections to 5 GHz flux for those sources are also given
in Table~2, and they range from 41 up to 980.

\section{The Model}
\label{sec:model}

We model the time evolution of an accretion disk under the radiation 
pressure instability. We adopt
the viscosity law using the geometrical mean between the gas and the
total pressure for the scaling of the stress, i.e.
$T_{\rm r \phi} \propto \alpha \, (P_{\rm gas}P_{\rm tot})^{1/2}$.
As was shown by Merloni
\& Nayakshin (2006), such a viscosity law nicely reproduces the effective
viscosity resulting from the magnetorotational instability, and describes
well the behaviour of the Galactic sources. 
This viscosity law is also favored for the black hole X-ray binaries 
(Done \& Davis 2008).

We are using the time-dependent numerical code developed by ourselves
and already presented in a number of works (Janiuk, Czerny \&
Siemiginowska 2002; Janiuk \& Czerny 2005; Czerny et al. 2008). The
initial conditions are determined by solving the energy balance,
$F_{\rm tot} = Q^{+}_{\rm visc} = Q^{-}_{\rm adv}+Q^{-}_{\rm rad}$, at
every disk radius $r$. Here, the total energy flux $F_{\rm tot}$
dissipated within the disk at a radius $r$ is calculated as:
\begin{equation}
F_{\rm tot} = {3 G M \dot M \over 8 \pi r^3} f(r)
\label{eq:ftot}
\end{equation}
\noindent
where the boundary condition is $f(r)=1-\sqrt{3\over r}$. The cooling and heating terms
in this vertically integrated (over the scale $H$) energy balance are:
\begin{equation}
Q^{+}_{\rm visc} = {3 \over 2}\alpha\Omega H \sqrt{P_{\rm gas}P_{\rm tot}}
\label{eq:heat}
\end{equation}
\noindent
where $\alpha$ is the viscosity parameter,
\begin{equation}
Q^{-}_{\rm rad}={ P_{\rm rad} c \over \tau}={ \sigma T^{4} \over 
\kappa \Sigma}
\label{eq:qrad}
\end{equation}
\noindent
and
\begin{equation}
Q^{-}_{\rm adv}=
 {2 r P q_{\rm adv} \over 3 \rho GM} F_{\rm tot}
\label{eq:fadv}
\end{equation}
\noindent
with $q_{\rm adv} \approx const \sim 1.0$.
The gas and radiation pressure are given by 
$P_{\rm gas}=k_{\rm B}/m_{\rm p}\rho T$ and $P_{\rm rad} = 4/3\sigma T^{4}$, 
$\tau$ is the optical depth, $\Sigma = \rho H$ is the gas column density, 
$c$, $k_{\rm B}$ and $\sigma$ are physical constants, and we adopt the electron 
scattering opacity $\kappa=0.34$ cm$^{2}$ g$^{-1}$. The hydrostatic balance 
equation gives ${P \over \rho} = \Omega^{2} H^{2}$, and the Keplerian angular 
velocity $\Omega$ is assumed.

We solve the basic evolutionary equations in 1-D and compute the thermal and 
viscous evolution of the accretion disk:
\begin{equation}
{\partial \Sigma \over \partial t}={1 \over r}{\partial \over \partial
r}(3 r^{1/2} {\partial \over \partial r}(r^{1/2} \nu \Sigma))
\end{equation}
\noindent
and
\begin{eqnarray}
{\partial T \over \partial t} + v_{\rm r}{\partial T \over \partial r}
= {T \over \Sigma}{4-3\beta \over 12-10.5\beta}
({\partial \Sigma \over
\partial t}+  v_{\rm r}{\partial \Sigma \over \partial r}) \\
\nonumber +{T\over P H}{1\over 12-10.5\beta} 
(Q^{+}-Q^{-}).
\end{eqnarray}
\noindent
Here  
\begin{equation}
v_{\rm r} = {3 \over \Sigma r^{1/2}} {\partial \over \partial r}
(\nu \Sigma r^{1/2})
\end{equation}
\noindent
is the radial velocity in the disk while
$\nu=(2 \sqrt{P_{\rm gas}P_{\rm tot}}\alpha)/(3\rho\Omega)$ 
is the kinematic viscosity and $\beta$ is the ratio of the gas pressure 
to the total pressure, $\beta = P_{\rm gas}/P $. 

The code also incorporates the option of a two-phase accretion, with
the mass exchange between the disk and a hot corona, which allows us
for performing quasi-2D computations. It can also include the effect
of the jet formation in parametric description. However, in the present
work, we concentrate only on the evolution of the disk, so the corona
and jet are neglected in order to keep the picture as simple as
possible. The models are thus parameterized by the black hole mass,
$M$, dimensionless accretion rate, $\dot m$, measured in the Eddington
units ($\dot M_{Edd} = 3.53\,(M/10^8 M_{\odot})\,M_{\odot}$\,yr$^{-1}$) 
and the viscosity parameter $\alpha$.

\section{Results}
\label{sec:results}

\begin{figure*}
\includegraphics[width=\columnwidth]{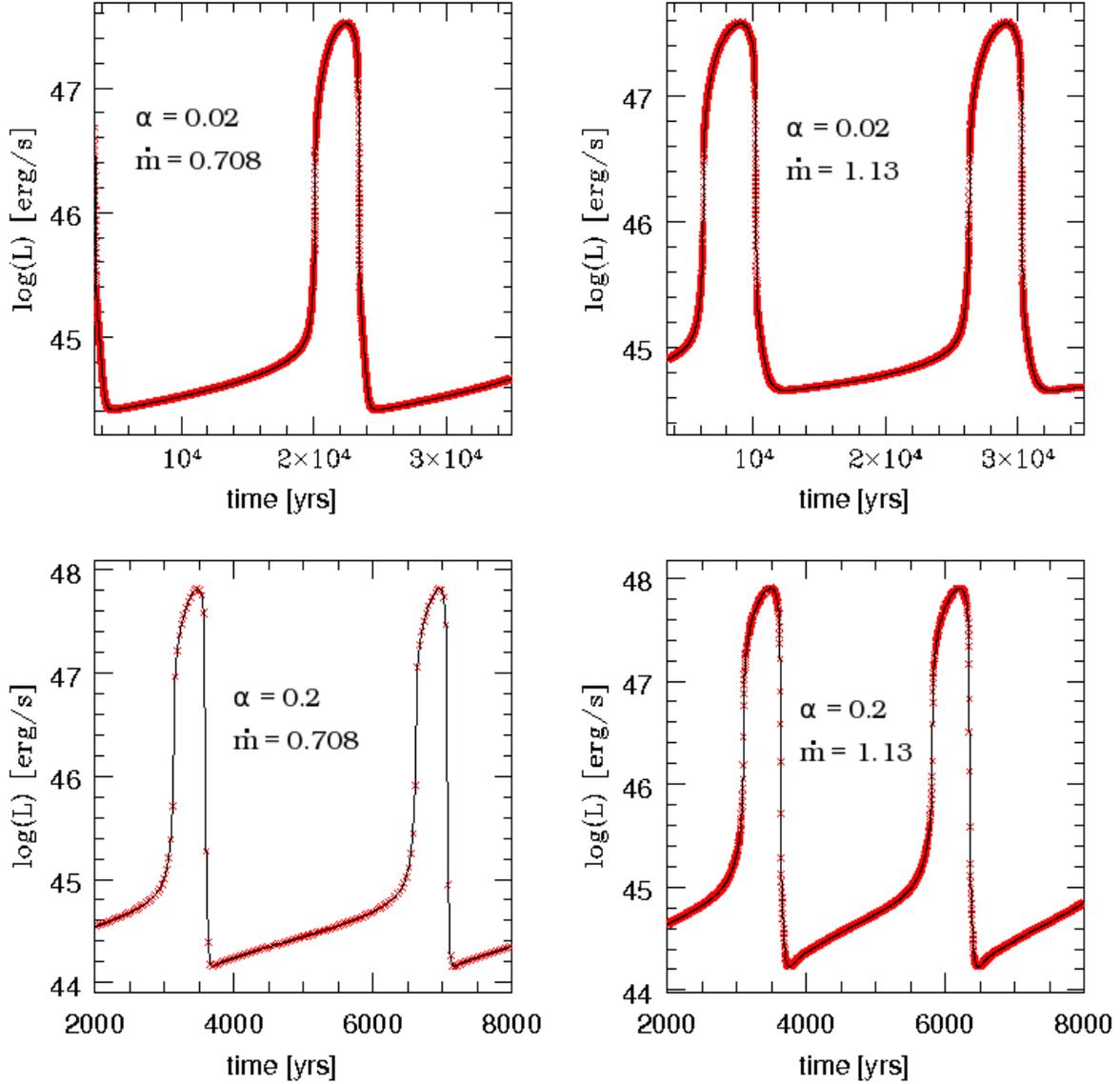}
\caption{Exemplary evolution of the disk bolometric luminosity under the 
radiation pressure instability ($T_{r \phi} \propto \alpha (P_{\rm
gas}P_{\rm tot})^{1/2}$) for two exemplary values of the accretion
rate $\dot m$ (in the Eddington units), and different viscosity
parameter $\alpha$. Black hole mass is taken as $M = 10^8\,M_{\odot}$.
\label{fig:lightcurve}}
\end{figure*}

\begin{figure*}
\includegraphics[width=\columnwidth]{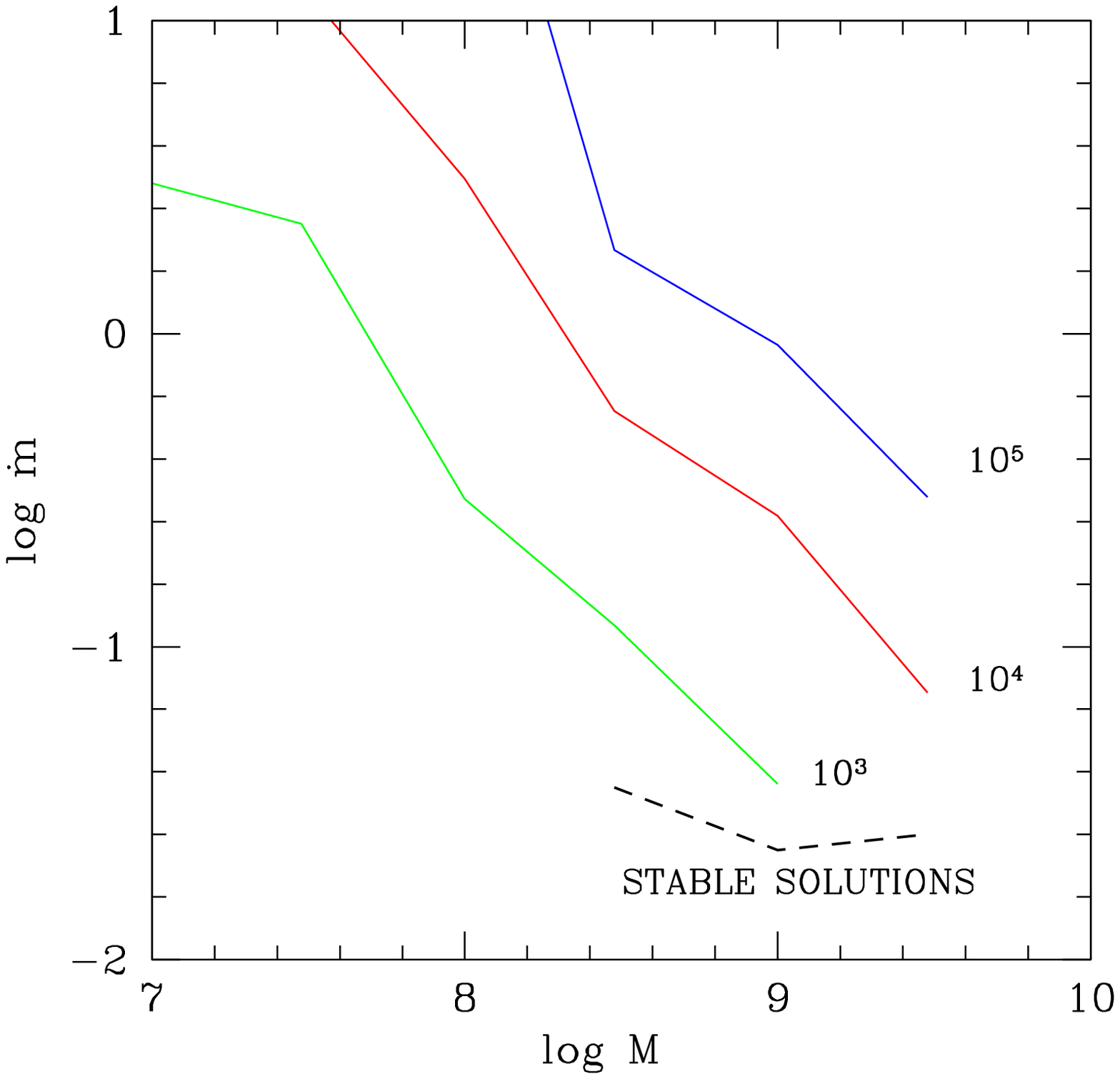}
\caption{Contour maps of the constant outburst duration time,
in the black hole mass vs. accretion rate (Eddington units) plane.
The outburst durations are given for each curve in years.
The viscosity parameter is taken as $\alpha=0.02$.
\label{fig:Tout02}}
\end{figure*}

\begin{figure*}
\includegraphics[width=\columnwidth]{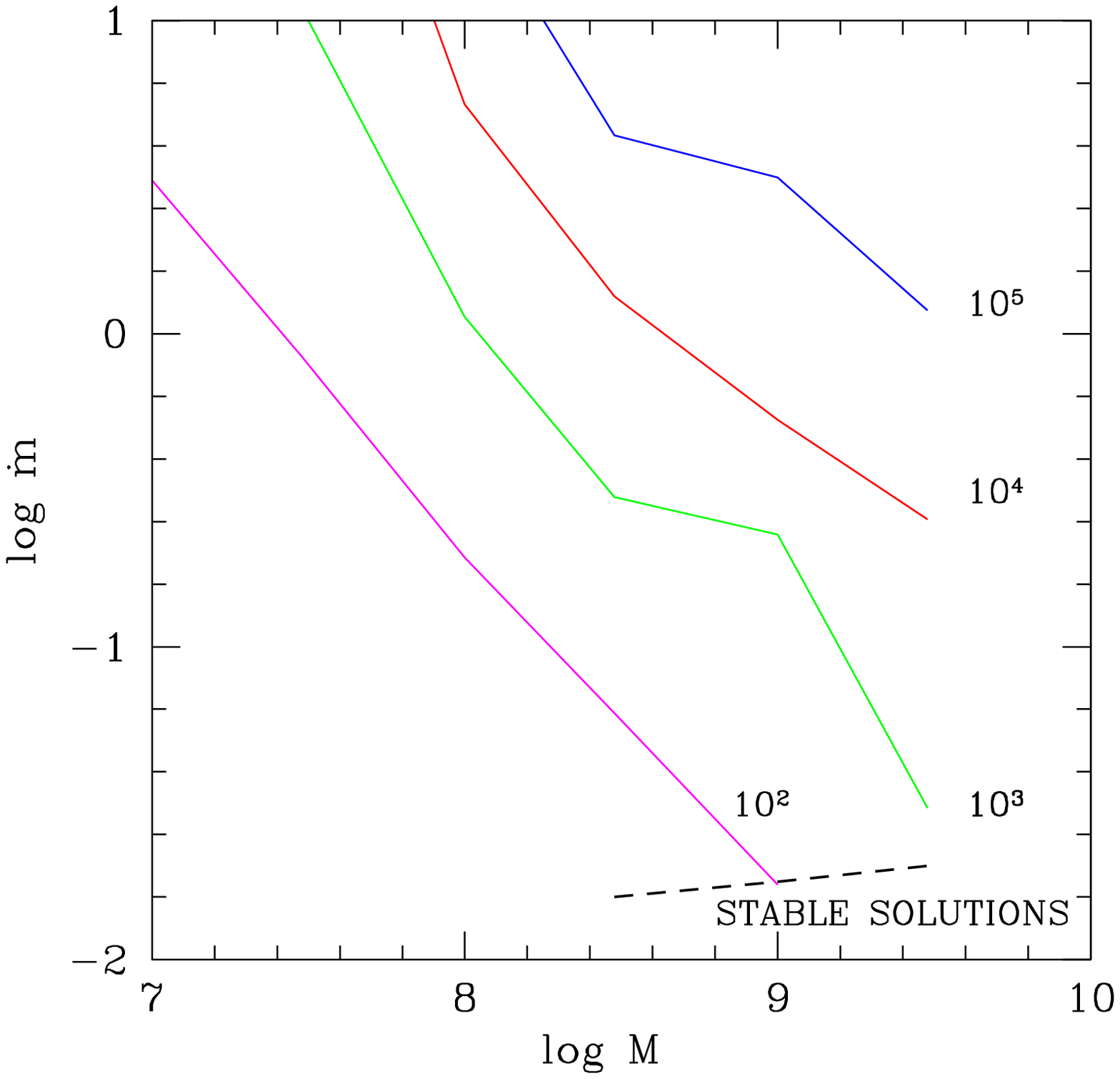}
\caption{Contour maps of the constant outburst duration time,
in the black hole mass vs. accretion rate (in Eddington units) plane.
The outburst durations are given for each curve in years.
The viscosity parameter is taken as $\alpha=0.2$.
\label{fig:Tout002}}
\end{figure*}

The time evolution of an accretion disk under the radiation pressure
instability is rather fast, in contrast with expectations based on the
partial ionization instability. The exemplary lightcurves representing
the evolution of the disk bolometric luminosity for a black hole
of mass $M = 10^{8}\,M_{\odot}$ are shown in Fig.~\ref{fig:lightcurve}.
We performed computations for a range of black hole masses likely to
be appropriate for GPS sources ($M= 10^7 - 3 \times 10^9\,M_{\odot}$). 
The expected durations of outbursts are shown in
Fig.~\ref{fig:Tout02} and Fig.~\ref{fig:Tout002} for the value of the
viscosity parameter $\alpha$ equal $0.2$ and $0.02$, correspondingly.
The duration of an outburst depends almost linearly on the black hole
mass, and strongly varies with the accretion rate. For sources close
to the Eddington limit expected outbursts last $10^4 - 10^5$\,yrs for
the smaller value of the viscosity parameter ($0.02$) and about ten times
shorter for the larger viscosity ($0.2$). Still shorter outbursts are
expected for sub-Eddington rates.

The overall range of the observed source ages ($200 -3 \times
10^5$\,yrs) is roughly consistent with the expected outburst duration
for the value of viscosity parameter $\alpha = 0.02$. The larger
viscosity parameter ($\alpha = 0.2$) requires super-Eddington
accretion rates to explain the duration of the oldest sources ($>10^5$
years).  Since we do not have the exact black hole mass determinations
for the sources in our sample (Table~\ref{tab1}) we cannot directly
locate them on the the contour maps shown in Figs.~\ref{fig:Tout02}
and ~\ref{fig:Tout002}. However, the lines of a constant time duration
of an outburst are almost parallel to the lines of a constant
bolometric luminosity for a source with an unknown black hole
mass. Therefore, for a given estimate of the source bolometric
luminosity we predict almost uniquely the duration of the outburst. If
in addition we relate the bolometric luminosity with the 5 GHz
monochromatic luminosity through a bolometric correction $K_{5GHz}$,
we obtain a relation:
\begin{eqnarray}
\label{eq:limit}
\log \left({T_{\rm burst} \over {\rm yr}}\right) \approx 1.25 \, \log 
\left({\nu L_{\nu} (5 GHz) \over {\rm erg/s}}\right) 
 + 0.38 \, \log \left({\alpha \over 0.02}\right) 
+ 1.25 \log K_{5GHz}- 53.6.
\end{eqnarray}
\noindent
The real age of the source cannot violate this limit, i.e. should be
shorter, and on average should correspond to half of the outburst
duration with some dispersion around this value.

\begin{figure*}
\includegraphics[width=\columnwidth]{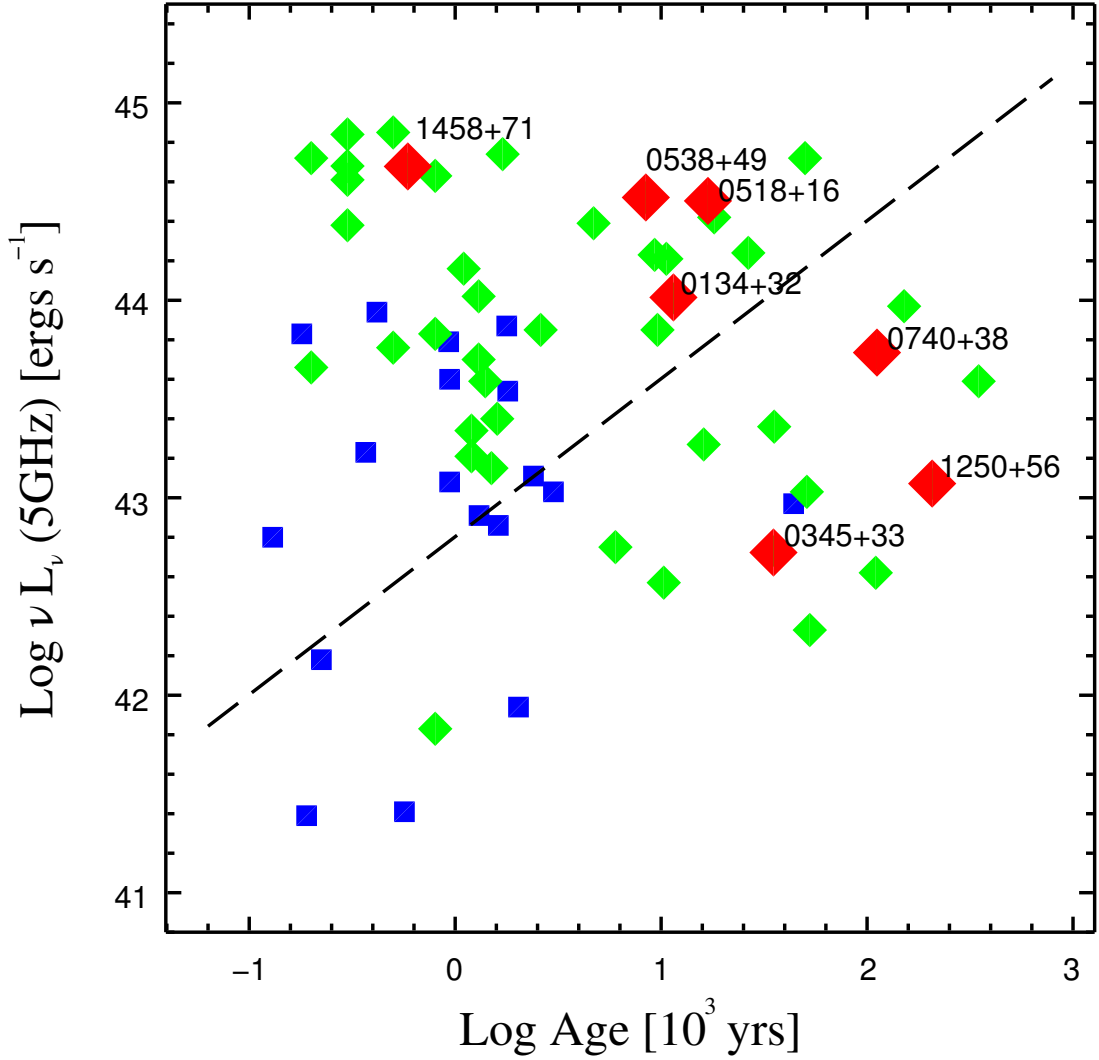}
\caption{Age vs. luminosity diagram for all the sources in the sample. 
The sources with measured black hole masses are shown by red large
diamonds. For those the younger sources show higher luminosity for the
same range of mass. More mass measurements are needed to test the
radio luminosity vs. age trends. Other sources with synchrotron age
determination are shown with green smaller diamonds, and sources with
kinematic age determination are shown with blue squares. The dashed
lines indicate the duration of an outburst (see Eq.8) for
$\alpha=0.02$ and the anticipated bolometric correction factor $300$.}
\label{fig:agelum}
\end{figure*}

In Fig.~\ref{fig:agelum}, we plot the lines corresponding to
Eq.~\ref{eq:limit} with $\alpha = 0.02$ and the anticipated bolometric
correction factor $300$ (see Sect.~\ref{sec:data}). On average, most
of the sources are located to the left of the line, as expected if the
age of the radio structure reflects directly the age of the central
engine. However, there are several outliers. Most of them are
relatively old, having their ages determined from the synchrotron
method (e.g., 1233+418). This method gives an upper limit to the age,
and thus the lifetimes of these, being about $\sim 10^5$\,yrs, may be
considerably overestimated. However, four sources with kinematically
measured age also show considerable departure to the right from the
line. The distance from the predicted line is the largest for
0035+227. Its radio spectrum is steep (e.g. Anton et
al. 2004). Perhaps the luminosity of the source is already dropping
down due to aging and therefore our standard relation between the
accretion rate in the model and the observed luminosity does not
apply.

Also, the sources should not be observed in the range where
instability does not operate. The exact boundary of the unstable
region is complicated and depends on the accretion rate. We can say
that the instability operates if $\dot{m} > 0.025$,
approximately. None of the sources in our sample with a measured mass
violate this limit. As argued in Section~\ref{sec:data}, we expect
that this criterium is also fulfilled on average, at least for the
majority of the GPS population.

\begin{figure*}
\includegraphics[width=\columnwidth]{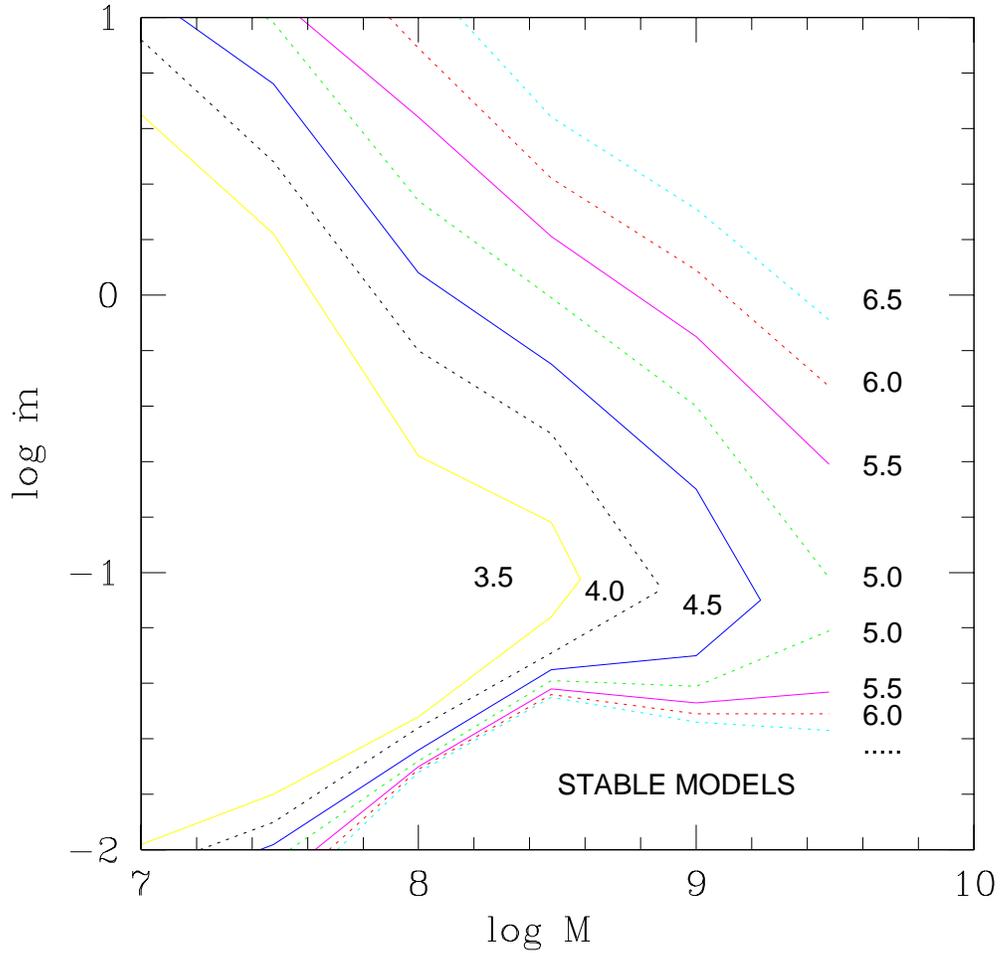}
\caption{Contour maps of the constant outburst separation time,
in the black hole mass vs. accretion rate (in the Eddington units). The
outburst durations are given for each curve, in $\log(T)$ [yrs].
The viscosity parameter is $\alpha=0.02$. A region of the parameter
space below the lowest curve shows shows stable model for accretion
rates below the threshold.}
\label{fig:Tquiet02}
\end{figure*}

\begin{figure*}
\includegraphics[width=\columnwidth]{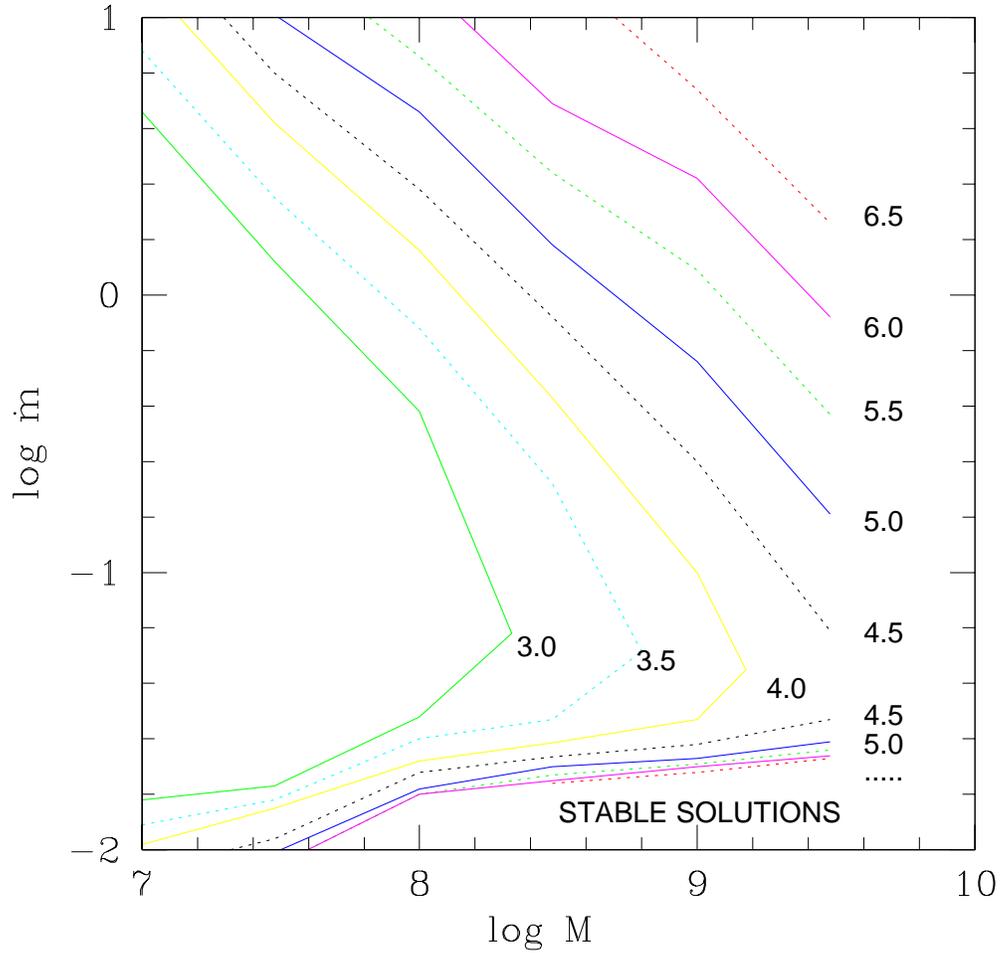}
\caption{Contour maps of the constant outburst separation time,
in the black hole mass vs. accretion rate (in the Eddington units).
The outburst durations are given for each curve, in $\log(T)$ [yrs].
The viscosity parameter is $\alpha=0.2$. A region of the parameter
space below the lowest curve shows shows stable models for accretion
rates below the threshold.}
\label{fig:Tquiet002}
\end{figure*}

In Fig.~\ref{fig:Tquiet02} and Fig.~\ref{fig:Tquiet002} we give the
outburst separations for lower and higher values of the viscosity
coefficient, respectively.  As the figure shows, the separation
between the outbursts is a complex non-monotonic function of the
accretion rate. For high accretion rates, the outburst separation
decreases with the accretion rate and increases with the black hole
mass. This trend is seen in Figs.~\ref{fig:Tquiet02}
and~\ref{fig:Tquiet002} as the upper branches of the constant time
separation lines.  This part of the plot is similar to the plots of
the outburst duration (Figs. \ref{fig:Tout02} and \ref{fig:Tout002}),
with the outburst separation being typically by a factor of 10
higher. The typical timescales are then consistent with the ones
postulated by Reynolds \& Begelman (1997). These timescales give the
limits for statistical studies of the galactic activity, and we devote
a part of the Discussion to this issue (see
Sect.~\ref{sec:evolution_diss}).
   
However, when the accretion rate approaches the value for which the instability 
ceases to exist, the trend with the accretion rate is reversed and the time 
separation goes to infinity. The limiting value of the accretion rate is specific 
for the given black hole mass. This is seen as the lower 
branches of the constant separation time lines. The parameter range (accretion 
rate and black hole mass) where this happens is relatively narrow. At still lower 
accretion rates, for a given mass, the disk is stable and there are no outbursts. 
Such a source should evolve in a continuous rather than intermittent way. 

\section{Discussion}
\label{sec:discussion}

We study the possibility that the observed short lifetimes of (some
of) GPS sources reflect the intermittent activity of the nucleus
caused by the radiation pressure instability in the accretion disk. We
model the disk outbursts using $T_{r \phi} \propto \alpha \, (P_{\rm
gas}P_{\rm tot})^{1/2}$ viscosity law, as previously done by Merloni
\& Nayakshin (2006) in the context of Galactic sources. This
parametric approach seems to be quite successful. For the smaller of
the two values of the considered viscosity parameter ($\alpha =
0.02$), supported by the previous studies of the AGN variability
(Siemiginowska \& Czerny 1989; Starling et al. 2004), the
theoretically obtained outburst durations for a range of black hole
masses and accretion rates are comparable to the ages of the radio
structures determined for our sample of 72 GPS/CSO objects.
Therefore, the radiation pressure instability mechanism offers an
interesting possibility for the intermittency of young radio sources,
as postulated by Reynolds \& Begelman (1997) to explain the apparent
overabundance of such. The time separations between outbursts obtained
from our model are also comparable to those requested by Reynolds \&
Begelman. However, our approach leaves several open questions/problems
which should be addressed in the future in order to establish the
theoretical basis for the intermittent character of jet activity. We
discuss these problems below.

\subsection{Radiation pressure instability}
\label{sec:instability_diss}

On the theoretical side, it is still unclear whether the radiation
pressure instability operates in accretion disks. If the disk
viscosity is parameterized as in the classical paper of Shakura \&
Sunyaev (1973), the domination by radiation pressure leads to the thermal
(Pringle et al. 1973) and viscous (Lightman \& Eardley 1974)
instability when the ratio of the gas pressure to the total pressure
drops below $0.4$ (Shakura \& Sunyaev 1976). If the viscosity scales with
the gas pressure, a stable solution for the disk structure is obtained 
(Sakimoto \& Coroniti 1981), and any intermediate scaling limits the parameter
range for the instability (with the scaling adopted in the present paper,
this limit is the gas pressure to the total pressure ratio of 3/11; see
Szuszkiewicz 1990). However, the viscosity mechanism that operates in
the accretion disks is connected with the magnetorotational instability
(Balbus \& Hawley 1991). Numerical simulations roughly support some
scaling of the effective torque with the pressure (linear for the gas
pressure dominated models, e.g. Hawley \& Krolik 2001, and with
$P_{\rm rad}^{0.7}$ scaling for the radiation pressure dominated models,
Hirose et al. 2008), although the issue of the global disk stability
is unclear. Older hydrodynamical simulations (e.g. Agol et al. 2001)
suggested that radiation pressure dominated disks are unstable while
in the most recent MHD simulations only a moderate variability was
found, without the global coherent outbursts (Turner 2004, Hirose et
al. 2008). However, these simulations still do not fully describe the
accretion process as they are limited to the shearing box
approach. 

Therefore, at this moment our parametric approach adopted to
compare the model predictions with the data seems to be the optimal
choice. In case of the regular periodic outbursts of the galactic source
GRS~1915+105, lasting from $\sim 100$ to $\sim 2000$ seconds (depending
on the source mean luminosity), this approach is successful (e.g.
Janiuk et al.  2002), particularly if the viscosity is parameterized as
$\alpha \, (P_{\rm gas}P_{\rm tot})^{1/2}$ (Merloni \& Nayakshin 2006). 
No other quantitative mechanism has been proposed to explain the observed 
behavior of this object. The frequently quoted evacuation of the inner 
disk due to a related outburst is rather unlikely (e.g. Czerny 2006), and
only the limit cycle mechanism (likely driven by the radiation pressure
instability) explains the absence of the direct transitions from the
state C to the state B in this black hole binary (Janiuk et al. 2002), 
as well as the observed QPO
oscillations (Misra et al. 2006). In addition, scaling the observed 
GRS~1915+105 timescale with the black hole mass by a factor $10^8$ gives
the outbursts durations from $300$\,yrs to $6000$\,yrs. This gives an 
additional, model-independent argument that an intermittency in AGN on 
the timescales of hundreds/thousands of years is quite likely to be of 
the similar origin indeed. 

\subsection{High and low luminosity states of accretion disk}
\label{sec:luminosity states}

The radiation pressure instability implies that an accretion disk
alternates between the two basic states which are thermally stable:
(1) an outburst, i.e. a high accretion rate state; and (2) a low
accretion rate state. During an outburst the disk is almost
stationary, with a roughly constant accretion rate at all radii, and
models developed for the stationary accretion disks apply to the
timescales shorter than the viscous timescale of hundreds/thousands of
years (e.g. Czerny 2006). A temporary accretion rate in this state is somewhat
higher than the Eddington rate, and higher than the external supply
rate to the outer parts of the disk.  The disk thermal stability
during the outburst is caused mainly by the advection of the part of
the energy to the black hole. A low state occurs between the outbursts
when the accretion rate falls rapidly and the disk becomes
significantly non-stationary - the local accretion rate decreases
inwards and the mass accumulates in the disk. The thermal stability in
this state is caused by a reduction of the radiation
pressure. Additionally, as the accretion rate becomes very low the
disk may become optically thin and/or replaced by an ADAF flow in its
innermost part (Janiuk et al. 2004). This effect was not included in
the computations since the issue of the cold disk evaporation and
disk/ADAF transition is a complex and open question (for a recent
review, see e.g. Narayan \& McClintock 2008). It probably does not
affect a global evolution although it would strongly influence the
radiation spectrum.

In our model we identified the high accretion rate state of the disk
with a high energy output of a jet. This is an assumption since the
jet/disk connection in Active Galaxies is not well
understood. However, a monotonic relation between the radio power of
the jet and the optical luminosity of the active nucleus (and thus
accretion rate) is suggested by the data for the entire population of
radio loud elliptical-hosted active galaxies (see e.g. Sikora et
al. 2007 and references therein). So even though a fraction of the
available accretion energy used to power a jet may vary between
objects with different Eddington ratios, the high accretion rate radio
loud sources generate more powerful jets than the low accretion rate
ones. That seems to be true regardless of the fact that the Eddington
ratio may not be the only parameter controlling the jet production
efficiency in AGN.  (e.g. Sikora et al. 2007, for the discussion on a
role of a black hole spin)

The radiation pressure instability should also operate in radio-quiet
sources. An outburst in this case would correspond to a typical state
of a bright radio-quiet quasar with a strong outflow while the low
state (between the outbursts) could be identified with sources
accreting at the Eddington ratio of about a fraction of a per
cent. Such low accretion sources are usually found among Low
Luminosity AGN. However, it is not easy to confirm that indeed the
bright AGN and (a fraction of) LL AGN are stages of the same major
strong activity episode. Observations of distant LL AGN are not
available, while the space density of bright nearby quasars is very
low, and the local LL AGN are most likely starved due to a shortage of
the accreting material. The selected radio-quiet samples are strongly
biased against the lower Eddington ratio sources, partially due to a
possible transition to an inner radiatively inefficient flow in this
state, as recently discussed by Hopkins et al.(2009). In other words,
the accretion disk intermittency presented in this paper, when applied
to radio quiet AGN may be related to the problem of ``missing AGN'' in
the context of the extragalactic X-ray background. In particular, as
noted by Hopkins et al. (2009), ``missing'' AGN needed to produce the
observed diffuse emission above 10~keV may not be heavily obscured,
but rather the ones accreting at lower rates. It is interesting to
speculate that these are in fact the AGN with their disks in the
quiescence, so in between the outbursts driven by the radiation
pressure instability.

Another manifestation of the radiation pressure instability would be
an observation of a fading quasar. However, the transition from the
high to low accretion rate state occurs on thermal timescale which is
of order of $\sim$350 years for a $10^8 M_{\odot}$ mass black hole,
and a factor of ten longer for a typical bright quasar hosting a black
hole mass of $10^9 M_{\odot}$.  On the other hand a lower mass Narrow
Line Seyfert 1 galaxies might be good candidates for monitoring the
effects of the radiation pressure instability. Some of them (although
most likely not all of them; see e.g. Mathur \& Grupe 2005, Nikolajuk
et al. 2009) are accreting at high Eddington ratios. Since black hole
masses in NLS1 galaxies are typically of order $10^6 - 10^7 M_{\odot}$
the activity decay timescales should be shorter, so NLS1 objects
should fade and transition to a LL AGN state in 4 - 40 years. A
outburst duration is by a factor 11 (for the viscosity parameter 0.02)
longer than the fading timescale. Therefore, one out of 11 objects in
a long monitored sample should change its spectral state from NLS1 to
S1 or LL AGN. An example of the NLS1 source disappearance was found by
Grupe et al. (2007), although the observational effect in this case
may be caused by a temporary obscuration as this particular source
seems to be brightening again on the timescales that are too short to
be explained by the state transitions (Grupe et al. 2008).

\subsection{Source evolution between the outbursts}
\label{sec:evolution_diss}

In the radiation pressure instability scenario, the outbursts repeat
regularly, every $10^4 - 10^6$ years, as long as there is a
considerable supply of accreting matter toward a black hole. If there
is one-to-one correspondence between the radio structure and the
activity epoch of a central engine, a ``young'' radio source would
just mean one more ongoing episode of the nuclear outburst. Meanwhile,
the previous phases of the jet activity should then manifest by the
presence of fossil radio lobes in addition to the compact (young)
ones, if only the timescale for fading of the latter structures is on
average longer than the characteristic quiescence timescale between
the activity epochs.  Also, some number of dying but compact radio
objects lacking active nuclei should be expected, depending on the
relative durations characterizing the outbursts, the quiescence
periods, and aging of dying/fossil radio lobes.  We note that about
$10\%$ of GPS objects (though mostly quasars) show evidences for a
faint extended radio structure in addition to the compact ones
(Stanghellini et al. 1998, 2005). Also, in some cases the traces of
complexity in compact radio morphologies are seen (Jeyakumar et
al. 2000). On the other hand, an extensive search for the evidence of
radio fossils in 374 radio sources with different ages done by
Sirothia et al. (2008) did not bring positive results. This may
indicate that the remnants of previous active phases are in general
rare, at least in the evolved/old objects. However, finding of such
remnants requires deep, low-frequency and high dynamic range radio
observations, and thus more analysis is needed before drawing the
robust conclusions.

But is the timescale for fading of compact radio lobes which stopped
to be supplied by the jet longer than the quiescence periods in
between of the accretion disk outbursts? If so, a significant number
of fossil radio structure should be indeed expected for GPS
sources. If not, the traces of the previous activity phase would be
erased before the new radio structure is born.  Note in this context
that Reynolds \& Begelman (1997) assumed that the radio source has an
initial outburst that lasts for $\sim 10^4$\,yrs and then the power
source switches off. The system continues its expansion (coasting
phase) that is is still pressure driven and the shocked shell remains
highly supersonic for the rest of the expansion period until $\sim
10^6$\,yrs. In their model, the radio emission fades while the
dynamical structure remains basically intact.  However, if the source
turns off at $\sim 10^3$~yrs, i.e. earlier than at the time considered
by Reynolds \& Begelman (1997), the high density galactic medium
should dramatically impact the source evolution quickly after the
power is turned off.

A typical advance speed of the CSO components is on the order of
$0.1\,c$ (Gugliucci et al. 2005), so during the active phase lasting
$t \sim 10^3$ yrs the ejecta cover the distance of $300$~pc. Further
pressure driven expansion proceeds until the accumulated jet energy
becomes comparable to the energy content of the heated
medium. Assuming the jet energy output rate of $L_{\rm j} \sim
10^{46}$\,erg\,s$^{-1}$, in agreement with the constraints presented
in Section~\ref{sec:data}, and the energy content of the heated medium
$E = R^{3}\,\rho\,v_{\rm s}^{2}$, with the number density $n = \rho /
m_{\rm p} \sim 0.1$\,cm$^{-3}$ and the sound speed $v_{\rm s} = (5 k T
/ 3 m_{\rm p})^{1/2} \sim 3.7 \times 10^7$\,cm\,s$^{-1}$ as
appropriate for the hot ($T \sim 10^7$\,K) gas in the central parts of
giant elliptical, we obtain the maximum size of the perturbed region
of $\sim 3$\,kpc for $E \sim L_{\rm j} \, t$.  Note that this is
consistent with the recent observations indicating a correlation
between the intensity of narrow emission lines (e.g. [OIII]~$\lambda
5007\AA$) and a linear size of a young radio source (see Labiano
2008). Because the NLR clouds are distributed within a few kpc from
the active nucleus, the overpressured expanding radio lobes with
comparable maximum sizes are expected to modify the line luminosity by
driving shocks into the ISM, as discussed in Labiano (2008).  The time
needed for the ejecta to reach this distance with the speed of
$0.1\,c$ is $3 \times 10^4$\,yrs. Later, the heated cocoon cools down
and recollapses (see e.g. Kaiser et al. 2000). The recollapse phase
happens with the sound speed of the heated material, and hence it
lasts for about $10^6-10^7$\,yrs. After that time no traces of the
past activity are expected. This is however much longer than the
outbursts separation times given in Figures 7 and 8 for the
anticipated black hole masses and accretion rates.

On the other hand, the recollapse timescale may be shorter if the jet
power is significantly lower than $L_{\rm j} \sim
10^{46}$\,erg\,s$^{-1}$. Also, high efficiency in radiating out the
jet energy in forms of high-energy photons, or non self-similar
evolution of radio cocoons, may play an analogous role. For example,
in the non self-similar evolutionary model for GPS radio galaxies
considered by Stawarz et al. (2008), the sideway expansion of the
compact ($\sim 1$\,kpc) lobes is expected to be lower than the advance
speed of the terminal hotspot by a factor of at least 3, and the
radiative jet efficiency is expected to exceed $10\%$ in the early
phases of the source evolution. Such effects can possibly result in
fact that the time separation between nuclear outbursts may be long
enough to erase the traces of the previous radio structure before a
new one is born, at least in some objects. An interesting question to
ask is how those other objects would look like on radio maps. Note in
this context that the new-born jet propagating not within the
unperturbed interstellar medium of the host elliptical but within the
fossil lobes is expected to display a very narrow radio cocoon and
very high ($> 0.1\,c$) advance velocity, resulting from the ballistic
propagation of the jet in a rarefied environment (Kaiser et al. 2000,
Stawarz 2004). Such a morphology would be called the ``core-jet'' one
rather than the ``classical double''. In other words, the source with
the quiescence periods shorter than the timescale for the collapse of
the fossil kpc-scale lobes would be classified as a core-jet dominated
GPS galaxy rather than CSO. It is typically assumed that the core-jet
dominated GPSes are in fact intrinsically similar to CSOs but only
viewed at smaller inclination.  However, the alternative explanation
we propose here is that the former ones represent rather the objects
with the outburst and quiescence timescales much shorter
($<10^5$\,yrs) than the latter ones.

Accordingly to the discussion presented above, we can expect that a
radio source powered by a short-lived outbursts of the central
activity is not be able to escape from the host galaxy. So it will be
confined within the host as it is observed in GPS and CSO sources.  In
order for the radio source to grow a large scale radio structure,
similar to the ones observed in FRI/FRII radio galaxies, the active
phase needs to last longer than, roughly, $10^4$\,yrs. This gives us a
limit on the accretion rates that are required for the formation of
the most extended radio sources. Based on Figures 5 and 6 and assuming
$10^4$\,yrs outburst durations, we estimate such a critical accretion
rate to be $0.5\,\dot{M}_{\rm Edd}$ for $\alpha=0.2$, and
$0.1\,\dot{M}_{\rm Edd}$ for $\alpha=0.02$.  Only few such powerful
outbursts would be needed to support the growth of a large radio
source beyond the host galaxy.

\subsection{Future applications}
\label{sec:applications_diss}

The model predicts a complex behaviour of the radio sources depending
on the black hole mass and disk accretion rate: continuous jet 
activity, intermittent compact sources, and intermittent sources with ejecta
leaving the host galaxy. Therefore, it will be interesting to consider
carefully statistics of the observed number of radio sources, their lifetimes
and sizes, in comparison to the predictions of this model. Perhaps, the
largest radio objects experience the most powerful and long-lasting
outbursts of the activity. For example, the outbursts detected in the {\it
Chandra} X-ray images of Perseus cluster last for $\sim 10^6$\,yrs with
a separation time of $\sim 10^7$\,yrs (Fabian et al. 2003; see McNamara
\& Nulsen 2007) for other examples of X-ray clusters). New measurements of 
the ages in CSO objects, needed for such comparison, are continuously
coming from VLBA\footnote{Two new sources from Nagai et al. (2008),
with both kinematic and synchrotron ages of $\sim 2000$ and $\sim
30$\,yrs, are not included in the present paper.}, and the statistics
of jet-related X-ray cavities in galaxy clusters is increasing. Future
surveys, like the on-going DEVOS (Deep Extragalactic VLBI-Optical
Survey; Mosoni et al. 2006) will bring even more objects that can be
analyzed in this respect.

\section*{Acknowledgments}

We thank Marek Sikora and Brandon Kelly for very helpful discussions.
We alse thank the anonymous referee for comments.

This research has made use of the NASA/IPAC Extragalactic Database
(NED) which is operated by the Jet Propulsion Laboratory, California
Institute of Technology, under contract with the National Aeronautics
and Space Administration.

This work was supported in part by grant 1P03D00829 of the Polish
State Committee for Scientific Research and the Polish Astroparticle
Network 621/E-78/BWSN-0068/2008.

This research is funded in part by NASA contract NAS8-39073. Partial
support for this work was provided by the Chandra grants GO2-3148A,
GO5-6113X and GO8-9125A-R.



{}


\begin{deluxetable}{lrccccccc}
\scriptsize
\small
\tablewidth{0pt}
\tablenum{1}
\tablecaption{GPS/CSO sample.}
\label{tab1}
\tablehead{
\colhead{name} &
\colhead{$z$} &
\colhead{age} &
\colhead{method} &
\colhead{$S_{\rm 5\,GHz}$} &
\colhead{$\log L_{\rm 5\,GHz}$} &
\colhead{type} &
\colhead{$\log(M/M_{\odot})$} &
\colhead{ref.}\\
 & & [$10^3$\,yrs] & & [mJy] & [erg\,s$^{-1}$] & & &
}
\startdata
0035+227 & 0.096 & 0.567 & kin & 247.0 & 41.41 & CSO  & $-$ &  \\
0108+388 & 0.668 & 0.417 & kin & 1325.0 & 43.94 & CSO  & $-$ &  \\
0710+439 & 0.518 & 0.93 & kin & 1629.0 & 43.79 & CSO  & $-$ &  \\
1031+567 & 0.4597 & 1.8 & kin & 1200.0 & 43.54 & CSO  & $-$ &   \\
1245+676 & 0.107 & 0.19 & kin & 188.0 & 41.39 & CSO  & $-$ &  5 \\
1358+624 & 0.238 & 2.4 & kin & 1838.0 & 43.11 & CSO  & $-$ &  12 \\
1404+286 & 0.076 & 0.224 & kin & 2404.0 & 42.18 & CSO  & $-$ &  6 \\
1843+356 & 0.763 & 0.18 & kin & 778.0 & 43.83 & CSO  & $-$ & 7 \\
1943+546 & 0.263 & 1.306 & kin & 942.0 & 42.91 & CSO  & $-$ & 1 \\
2021+614 & 0.227 & 0.368 & kin & 2725.0 & 43.23 & CSO  & $-$ & 8 \\
2352+495 & 0.237 & 3.0 & kin & 1538.0 & 43.03 & CSO  & $-$ & 6 \\
B0840+424 & 0.35 & 44.0 & kin & 586.0 & 42.97 & CSO  & $-$ & 13 \\
J2203+1007 & 0.9 & 0.94 & kin & 316.0 & 43.60 & CSO  & $-$ &  \\
J1111+1955 & 0.299 & 1.62 & kin & 642.0 & 42.86 & CSO  & $-$ &  \\
1414+4554 & 0.186 & 2.03 & kin & 209.0 & 41.94 & CSO  & $-$ & 10 \\
J1415+1320 & 0.246 & 0.13 & kin & 839.0 & 42.80 & CSO  & $-$ & 10 \\
J1734+0926 & 0.735 & 1.78 & kin & 909.0 & 43.87 & CSO  & $-$ & 10 \\
J1915+6548 & 0.486 & 0.94 & kin & 367.0 & 43.08 & CSO  & $-$ & 11 \\
0221+67 & 0.31 & 51.0 & syn & 870.0 & 43.03 & GPS/CSS a & $-$ & 14 \\
0404+76 & 0.6 & 1.1 & syn & 2820.0 & 44.16 & GPS/CSS a & $-$ & 14 \\
0740+38 & 1.06 & 113.2 & syn & 304.0 & 43.73 & GPS/CSS a & 9.51 & 14 \\
1005+07 & 0.88 & 18.1 & syn & 2210.0 & 44.42 & GPS/CSS a & $-$ & 14 \\
1019+22 & 1.62 & 9.3 & syn & 391.0 & 44.23 & GPS/CSS a & $-$ & 14 \\
1203+64 & 0.37 & 35.4 & syn & 1279.0 & 43.36 & GPS/CSS a & $-$ & 14 \\
1250+56 & 0.32 & 204.3 & syn & 883.0 & 43.07 & GPS/CSS a & 8.31 & 14 \\
1323+32 & 0.37 & 1.4 & syn & 2153.0 & 43.59 & GPS/CSS a & $-$ & 14 \\
1416+06 & 1.44 & 50.0 & syn & 1550.0 & 44.72 & GPS/CSS a & $-$ & 14 \\
1443+77 & 0.27 & 110.2 & syn & 457.0 & 42.62 & GPS/CSS a & $-$ & 14 \\
1447+77 & 1.13 & 151.3 & syn & 457.0 & 43.97 & GPS/CSS a & $-$ & 14 \\
1607+26 & 0.47 & 0.5 & syn & 1908.0 & 43.76 & GPS/CSS a & $-$ & 14 \\
2248+71 & 1.84 & 26.5 & syn & 306.0 & 44.24 & GPS/CSS a & $-$ & 14 \\
2252+12 & 0.54 & 348.2 & syn & 935.0 & 43.59 & GPS/CSS a & $-$ & 14 \\
2342+82 & 0.74 & 1.3 & syn & 1280.0 & 44.02 & GPS/CSS a & $-$ & 14 \\
0810+460B & 0.33 & 10.3 & syn & 266.0 & 42.57 & GPS/CSS a & $-$ & 14 \\
1025+390B & 0.361 & 6.0 & syn & 332.0 & 42.75 & GPS/CSS a? & $-$ & 14 \\
1233+418 & 0.25 & 52.7 & syn & 276.0 & 42.33 & GPS/CSS a? & $-$ & 14 \\
0127+23 & 1.46 & 0.3 & syn & 1167.0 & 44.61 & GPS/CSS b & $-$ & 14 \\
0134+32 & 0.37 & 11.3 & syn & 5727.0 & 44.01 & GPS/CSS b & 8.74 & 14 \\
0138+13 & 0.62 & 1.3 & syn & 903.0 & 43.70 & GPS/CSS b & $-$ &14 \\
0223+34 & 1.0 & 0.3 & syn & 1515.0 & 44.38 & GPS/CSS b & $-$ & 14 \\
0316+16 & 1.0 & 0.3 & syn & 3032.0 & 44.68 & GPS/CSS b & $-$ & 14 \\
0428+20 & 0.22 & 1.2 & syn & 2764.0 & 43.21 & GPS/CSS b & $-$ & 14 \\
0429+41 & 1.02 & 0.5 & syn & 4301.0 & 44.85 & GPS/CSS b & $-$ & 14 \\
0518+16 & 0.76 & 16.7 & syn & 3652.0 & 44.50 & GPS/CSS b & 8.53 & 14 \\
0538+49 & 0.55 & 8.3 & syn & 7637.0 & 44.52 & GPS/CSS b & 9.58 & 14 \\
0758+14 & 1.2 & 10.6 & syn & 696.0 & 44.21 & GPS/CSS b & $-$ & 14 \\
1225+36 & 1.98 & 0.2 & syn & 812.0 & 44.72 & GPS/CSS b & $-$ & 14 \\
1328+30 & 0.85 & 0.3 & syn & 6191.0 & 44.84 & GPS/CSS b & $-$ & 14 \\
1328+25 & 1.06 & 1.7 & syn & 3068.0 & 44.74 & GPS/CSS b & $-$ & 14 \\
1358+62 & 0.43 & 0.2 & syn & 1838.0 & 43.66 & GPS/CSS b & $-$ & 14 \\
1458+71 & 0.9 & 0.6 & syn & 3733.0 & 44.67 & GPS/CSS b & 8.98  & 14 \\
1517+20 & 0.75 & 9.6 & syn & 841.0 & 43.85 & GPS/CSS b & $-$ & 14 \\
1819+39 & 0.4 & 1.6 & syn & 1179.0 & 43.40 & GPS/CSS b & $-$ & 14 \\
1829+29 & 0.6 & 0.8 & syn & 1311.0 & 43.83 & GPS/CSS b & $-$ & 14 \\
2249+18 & 1.76 & 0.8 & syn & 822.0 & 44.63 & GPS/CSS b & $-$ & 14 \\
0345+33 & 0.24 & 35.8 & syn & 762.0 & 42.73 & GPS/CSS ? & 7.75 & 14 \\
0809+404 & 0.55 & 16.1 & syn & 439.0 & 43.27 & GPS/CSS ? & $-$ & 14 \\
1128+455 & 0.4 & 1.5 & syn & 659.0 & 43.15 & GPS/CSS ? & $-$ & 14 \\
1159+395 & 2.37 & 4.7 & syn & 264.0 & 44.39 & GPS/CSS ? & $-$ & 14 \\
1225+442 & 0.22 & 0.8 & syn & 114 & 41.83 & GPS/CSS ? & $-$ & 14 \\
1242+410 & 0.811 & 2.6 & syn & 715.0 & 43.85 & GPS/CSS ? & $-$ & 14 \\
1350+432 & 2.149 & 1.2 & syn & 28.6 & 43.34 & GPS/CSS ? & $-$ & 14 \\
J0000+4054 & $-$ & 0.28 & kin & 676.0 & $-$ & CSO ? & $-$ &  \\
J1826+1831 & $-$ & 0.38 & kin & 434.0 & $-$  & CSO ? & $-$ & 10 \\
J0003+4807 & $-$ & 0.34 & kin & 197.0 & $-$  & CSO ? & $-$ & 10 \\
J0204+0903 & $-$ & 0.24 & kin & 715.0 & $-$  & CSO ? & $-$ & 10 \\
J0427+4133 & $-$ & 0.02 & kin & 723.0 & $-$  & CSO ? & $-$ & 10 \\
J0620+2102 & $-$ & 2.06 & kin & 442.0 & $-$ & CSO ? & $-$ & 10 \\
J0754+5324 & $-$ & 2.22 & kin & 286.0 & $-$ & CSO ? & $-$ & 10 \\
J1143+1834 & $-$ & 0.69 & kin & 279.0 & $-$ & CSO ? & $-$ & 10 \\
B0147+400 & $-$ & 6.2 & kin & 254.0 & $-$ & CSO ? & $-$ & 13 \\
1153+31 & $-$ & 5.4 & syn & 1.56 & $-$ & GPS/CSS a & $-$ & 14 \\
\enddata
\tablecomments{
References: 
[1] Polatidis \& Conway (2003);
[2] Owsianik et al (1998);
[3] Owsianik \& Conway (1998);
[4] Taylor et al (2000);
[5] Marecki et al (2003);
[6] Stanghellini et al (2002);
[7] Polatidis (2001);
[8] Tschager et al (2000);
[9] Polatidis et al (2003);
[10] Gugliucci et al (2005);
[11] Gugliucci et al (2007);
[12] Vink et al (2006);
[13] Orienti et al (2007);
[14] Murgia et al. (1999).
Black hole masses are taken from Woo et al. (2002) and Liu et al. (2006). 
Different types/classes of sources as explained in Section~\ref{sec:data}.
Ages of the sources with no redshift were estimated assuming $z = 0.5$ (Gugliucci et al. 2007).}
\end{deluxetable}

\begin{deluxetable}{cccccccc}
\scriptsize
\small
\tablewidth{0pt}
\tablenum{2}
\tablecaption{GPS/CSO sources with given black hole masses.}
\label{tab2}
\tablehead{
\colhead{name} &
\colhead{$\log M/M_{\odot}$} &
\colhead{$\log L_{\rm Edd}^a$} &
\colhead{$\log S_B^b$} &
\colhead{$\log L_B^c$} &
\colhead{$\log L_{\rm bol}^d$} &
\colhead{$\dot{m}^e$} &
\colhead{$K_{5GHz}^f$}\\
& & [erg\,s$^{-1}$] & [erg\,s$^{-1}$\,Hz$^{-1}$] & [erg\,s${^{-1}}$] & [erg\,s$^{-1}$] & & 
}
\startdata
0740+38 & 9.51 & 47.62 & 30.76 & 45.72 & 46.72 &  0.12 & 980\\
1250+56 & 8.31 & 46.42 & 29.85 & 44.74 & 45.74 &  0.21 & 470\\
0134+32 & 8.74 & 46.85 & 30.59 & 45.48 & 46.48 &  0.43 & 300\\
0518+16 & 8.53 & 46.64 & 30.88 & 45.82 & 46.82 &  1.49 & 290\\
0538+49 & 9.58 & 47.69 & 30.21 & 45.13 & 46.13 &  0.03 &  41\\
1458+71 & 8.98 & 47.09 & 30.89 & 45.84 & 46.84 &  0.56 & 150\\
\enddata
\tablecomments{
$^a$ Eddington luminosity;\\
$^b$ $B$-band (4400\AA) flux from Netzer et al. (1996);\\
$^c$ $B$-band (4400\AA) luminosity;\\
$^d$ Bolometric disk luminosity $L_{\rm bol} = 10 \times L_B$;\\
$^e$ Accretion rate in the Eddington units;\\
$^f$ Bolometric correction to 5 GHz monochromatic luminosity.
}
\end{deluxetable}

\label{lastpage}
\end{document}